




\global\newcount\ftno \global\ftno=1
\def\foot#1{{\baselineskip=14.4pt plus 0.3pt\footnote{$^{\the\ftno}$}{#1}}%
\global\advance\ftno by1}

\newcount\notenumber \notenumber=1
\def\note#1{\footnote{$^{\the\notenumber}$}{#1}%
\global\advance\notenumber by 1}

\voffset 1cm
\vglue 3cm
\centerline {\bf Dynamical Symmetry Breaking In A Nambu-Jona-Lasinio}
\centerline {\bf Model And The Vacuum Structure}
\vskip 2cm
\centerline {E. A. Dudas}
\centerline {Laboratoire de Physique Th\'eorique et Hautes
Energies\footnote*{Laboratoire associ\'e au Centre National de la
Recherche Scientifique}}
\centerline {Universit\'e de Paris-Sud, B\^at. 211, 91405 Orsay, France}
\vskip 2cm
\centerline {ABSTRACT}
\vskip 1cm
The vacuum structure for a Nambu-Jona-Lasinio type model is studied
using the effective potential approach. The relevant degrees of freedom
are taken to be two different sets of static, auxiliary fields with
different symmetry properties, related by Fierz transformations in the
action. The results are compared with the analysis of the
Dyson-Schwinger equations in the one-loop approximation.

\vfill
LPTHE 93/    (Juillet 1993)
\vfill
\eject
\noindent {\hbox { \quad {\bf 1. Introduction}}}
\vskip .5cm
   The idea of dynamical symmetry breaking in field theory [1] was recently
reconsidered in connection with the possibility of having a composite Higgs
particle [2]. The symmetry breaking is triggered by a top-antitop
condensation mechanism of the Nambu-Jona-Lasinio (NJL) type and the model
reproduces the usual low energy phenomenology.
   In dealing with NJL type interactions two different approaches, usually
considered equivalent, are employed. The first is the Dyson-Schwinger
(DS) equations for the propagators with possible nontrivial solutions
viewed as a signal of symmetry breaking and the appeareance of bound states.
The second consists in rewriting the interaction Lagrangian with the aid
of static, scalar auxiliary fields. Working out the corresponding effective
potential and minimizing it the symmetry breaking can be investigated [3].
   In the present paper a simple NJL type model is considered. Two
different sets of auxiliary fields with different symmetry properties
are introduced. The one-loop effective potential analysis gives different
results in the two cases and the real minimum is given by the
condensation of the scalar field constraints with the higher symmetry.
The naive analysis of the DS equations in the one-loop approximation
gives a third different answer. An explanation in the light of the
Fierz transformation is given and it is argued that the real vacuum
cannot be obtained in this model using the DS equations.
\vskip 1cm
\noindent {\hbox {\quad {\bf 2. The model, symmetries and the
Dyson-Schwinger equations}}}
\vskip .5cm
The model we will deal with contains two $SU(2)$ isodoublets $\psi_1$ and
$\psi_2$
and is described by the following Lagrangian
$$
{\cal L} = \bar \psi_1 i {\rlap /\partial} \psi_1 + \bar \psi_2 i  {\rlap
/\partial}
\psi_2\ +\ G(\bar \psi_1 {\tau \over 2}^a \psi_2) (\bar \psi_2 {\tau \over 2}^a
\psi_1)\hskip .5cm .\eqno(1)
 $$
In eq.(1) $\tau^a$ are the Pauli matrices and $G$ is a positive coupling. The
symmetries of the model can be summarized as follows :
$$\matrix{
&i)\quad SU(2)  \hskip 4cm & \psi_1 &\rightarrow &e^{{i \over 2} \theta^a
\tau^a}
\psi_1\hskip 3cm\cr
& \hskip 4cm& \psi_2 &\rightarrow &e^{{i \over 2} \theta^a \tau^a}
\psi_2 \hskip .5cm ,\hskip 2.5 cm\cr \vbox to 1cm{}
&ii)\quad U(1)_1  \hskip 4cm  &\psi_1 &\rightarrow &e^{i \alpha} \psi_1\hskip
3cm\cr
& \hskip 4cm &\psi_2 &\rightarrow &\psi_2\hskip .5cm ,\hskip 2.5cm\cr\vbox to
1cm{}
&iii)\quad U(1)_2 \hskip 4 cm & \psi_1 &\rightarrow & \psi_1\hskip 3cm\cr
&   \hskip 4cm &\psi_2 &\rightarrow &e^{i\beta} \psi_2\hskip .5cm
,\hskip 2.5cm\cr\vbox to 1cm{}
&iv)\quad U(1)_A   \hskip 4cm & \psi_1 &\rightarrow &e^{i \alpha \gamma_5}
\psi_1\hskip 3cm\cr
 &  \hskip 4cm &\psi_2 &\rightarrow &e^{-i\alpha \gamma_5}
\psi_2J\hskip .5cm .\hskip 2.5cm\cr
}$$
Due to the attractive four-fermion interaction we can have bound-state
formation
and dynamical breakdown of some symmetries by a Nambu-Jona-Lasinio type
mechanism [1]. There are various possibilities for the configurations of the
condensates
$$\matrix{
 a) \hfill \quad < {\bar \psi}_1 \psi_1 >\hfill &\not= 0\quad ,\quad
U(1)_A\hfill\cr
 b)\hfill\quad < {\bar \psi}_2 \psi_2 > \hfill&\not= 0\quad , \quad
U(1)_A\hfill\cr
 c)\hfill\quad < {\bar \psi}_2 \psi_1 > \hfill&\not= 0\quad , \quad
U(1)_{1-2}\hfill\cr
d) \hfill\quad < {\bar \psi}_1 {\tau \over 2}^a \psi_1 >\hfill &\not= 0\quad ,
\quad
SU(2) \times U(1)_A\hfill\cr e) \hfill\quad < {\bar \psi}_2 {\tau \over 2}^a
\psi_2 >
\hfill&\not= 0 \quad , \quad SU(2) \times U(1)_A\hfill\cr
f)\hfill\quad < {\bar \psi}_1 {\tau \over 2}^a \psi_2 >\hfill&\not= 0
\quad,\quad SU(2) \times
U(1)_{1-2}\hskip .5cm .\hfill\cr
}$$
\hskip.5cm
The symmetries specified in the right in the above formulae are those which are
broken by the appearance of the corresponding condensate and $U(1)_{1-2}$
means the combined $U(1)_1 \times U(1)_2$ transformations with $\beta = -
\alpha$.

In order to study the dynamical symmetry breaking we will use in a first
approach
the Dyson-Schwinger equations in the one-loop approximation.

Using the notation
$$
S^{-1}(p) = \rlap /p - \Sigma \hskip .5cm ,\eqno(2)
$$
where $\Sigma = \Sigma_0 + \tau_a \Sigma_a$ , we obtain
$$
\Sigma^{(1)}_{ij} = {iG \over 2} \int {d^4 k \over (2\pi)^4} \biggl[
\delta_{ij}
S^{(2)}_{\ell \ell} (k) - {1 \over 2} S^{(2)}_{ij} (k) \biggr]\hskip
.5cm ,
$$
$$
\Sigma^{(2)}_{ij} (p) = {iG \over 2} \int {d^4 k \over (2\pi)^4} \biggl[
\delta_{ij}
S^{(1)}_{\ell \ell} (k) - {1 \over 2} S^{(1)}_{ij} (k) \biggr]\hskip
.5cm . \eqno(3)
$$
In eq.(3) the indices (1) and (2) stand for the first and the second $SU(2)$
isodoublet. Making the Wick rotation and writing them explicitly we find the
following equations
$$\eqalign{
4 \Sigma^{(1)}_0 &= -3G \int {d^4k \over (2\pi)^4}\quad { k^2 +
\Sigma^{(2)^2}_0
- \buildrel \rightarrow \over \Sigma^{(2)^2} \over
\biggl(k^2 + \Sigma^{(2)^2} + \buildrel \rightarrow \over
\Sigma^{(2)^2}\biggr)^2 -
4 \Sigma^{(2)^2}_0 \buildrel \rightarrow \over \Sigma^{(2)^2}}
\quad \Sigma^{(2)}_0\hskip .5cm ,\cr
4 \Sigma^{(2)}_0 &= -3G \int {d^4k \over (2\pi)^4}\quad { k^2 +
\Sigma^{(1)^2}_0
- \buildrel \rightarrow \over \Sigma^{(1)^2} \over
\biggl(k^2 + \Sigma^{(1)^2} + \buildrel \rightarrow \over
\Sigma^{(1)^2}\biggr)^2 -
4 \Sigma^{(1)^2}_0 \buildrel \rightarrow \over \Sigma^{(1)^2}}
\quad \Sigma^{(1)}_0\hskip .5cm ,\cr
4 \Sigma^{(1)}_a &=  G \int {d^4k \over (2\pi)^4}\quad { k^2 - \Sigma^{(2)^2}_0
+ \buildrel \rightarrow \over \Sigma^{(2)^2} \over
\biggl(k^2 + \Sigma^{(2)^2}_0 + \buildrel \rightarrow \over
\Sigma^{(2)^2}\biggr)^2
- 4 \Sigma^{(2)^2}_0 \buildrel \rightarrow \over \Sigma^{(2)^2}}
\quad \Sigma^{(2)}_a\hskip .5cm ,\cr
4 \Sigma^{(2)}_a &=  G \int {d^4k \over (2\pi)^4}\quad { k^2 - \Sigma^{(1)^2}_0
+ \buildrel \rightarrow \over \Sigma^{(1)^2} \over
\biggl(k^2 + \Sigma^{(1)^2}_0 + \buildrel \rightarrow \over
\Sigma^{(1)^2}\biggr)^2
- 4 \Sigma^{(1)^2}_0 \buildrel \rightarrow \over \Sigma^{(1)^2}}
\quad \Sigma^{(1)}_a\hskip .5cm .\cr
}\eqno(4)$$
We are interested in the possible nontrivial solutions of eqs.(4). These are
given by
$$\eqalignno{
i)\hfill\hskip 2cm \Sigma^{(1)}_a &= \Sigma^{(2)}_a = 0\hskip .5cm ,\hfill\cr
 \Sigma^{(1)}_0 &=\ -\ \Sigma^{(2)}_0\hskip .5cm ,\hfill\cr
&\qquad{4 \over 3G} = \int {d^4k \over (2\pi)^4} {1 \over k^2 +
\Sigma^2_0}\hskip .5cm .\hskip 4.5cm\hfill &(5)\cr }$$
$$\eqalign{
ii)\hfill\hskip 2cm \Sigma^{(1)}_0 &= \Sigma^{(2)}_0 = 0\hskip .5cm ,\hfill\cr
 \Sigma^{(1)}_a &=\ +\ \Sigma^{(2)}_a\hskip .5cm ,\hfill\cr
&\qquad{4 \over G} = \int {d^4k \over (2\pi)^4} {1 \over k^2 + \buildrel
\rightarrow
\over \Sigma^2 }\hskip .5cm .\hskip 4.5cm\hfill \cr
}$$
The first solution correspond to the appearence of two anti alligned
condensates $<
\bar \psi_1 \psi_1 > = - < \bar \psi_2 \psi_2 >$ and the second to the
configuration
$<{\bar \psi}_1 {\tau \over 2}^a \psi_1 > = + <{\bar \psi}_2 {\tau \over 2}^a
\psi_2 >$. Before
trying to interpret the result we will use a second approach to the vacuum
structure, the effective potential [4].
\vskip 1cm
\noindent {\hbox {\quad {\bf 3. The effective potential approach}}}
\vskip .5cm
The idea is to write the interaction Lagrangian in two different ways using the
Fierz transformations. In each case we will introduce a set of static,
auxiliary
fields and construct the one-loop effective potential in the standard way [3].
Finally we will write the saddle point equations and compute the vacuum energy
in
each case.

The Fierz transformation allows us to rewrite the interaction Lagrangian as
follows
$$
{\cal L}_{int} = G \biggl( \bar \psi_1 {\tau \over 2}^a \psi_2 \biggr ) \biggl
(\bar
\psi_2 {\tau \over 2}^a \psi_1 \biggr) = - {G \over 4} \biggl [ (\bar \psi_1
\psi_2)
(\bar \psi_2 \psi_1) + 2 (\bar \psi_{1\alpha} \psi_{1\beta}) (\bar \psi_{2\beta
}
\psi_{2\alpha }) \biggr ]\hskip .5cm . \eqno(6)
 $$

Then we can introduce two different sets of auxiliary fields and write the
lagrangian in two equivalent ways :
$$\eqalign{
{\cal L} &= \bar \psi_1 i {\rlap /\partial} \psi_1 + \bar \psi_2 i  {\rlap
/\partial}
\psi_2\ +\ \varphi ^a \bar \psi_1 {\tau \over 2}^a \psi_2 +\ \varphi^{a+}  \bar
\psi_2 {\tau \over 2}^a \psi_1 - {1 \over G} \varphi^{a+} \varphi^a \cr
&= \bar \psi_1 i {\rlap /\partial} \psi_1 + \bar \psi_2 i  {\rlap /\partial}
\psi_2\ + \bar \psi_1\psi_2 \phi_1 + \bar
\psi_2 \psi_1\phi_1 - {4 \over G} \phi^+_1 \phi_1 + \bar \psi_{1\alpha}
\psi_{1\beta} \phi_2^{\alpha\beta} +\cr
&+ \bar \psi_{2\alpha } \psi_{2\beta} \phi_3^{\alpha\beta} +{2 \over G}
\phi_2^{\alpha\beta} \phi_3^{\beta\alpha}\hskip .5cm .}\eqno(7)
 $$

In the above expressions $\alpha$ and $\beta$ are Lorentz indices. The
auxiliary
fields are defined by their fields equations
$$\eqalign{
\varphi^a &= {1 \over G} \bar \psi_2 {\tau^a \over 2} \psi_1\hskip .5cm ,\cr
\phi_1 &= -{G \over 4} \bar \psi_2 \psi_1\hskip .5cm ,\cr
\phi^{\beta \alpha}_2 &= - {G \over 2} \bar \psi_2^{\alpha}
\psi_2^{\beta}\hskip .5cm ,\cr
\phi^{\beta \alpha}_3 &= - {G \over 2} \bar \psi_1^{\alpha}
\psi_1^{\beta}\hskip .5cm .\cr
}\eqno(8)
$$
With the first set of constraints $\varphi^a$ the one-loop effective potential
is
$$
V_{ef} (\varphi^a) = {1 \over G} \varphi^{a^+} \varphi^a - 2 \int {d^4 p \over
(2\pi)^4} \ell n \biggl [ (p^2 + {1 \over 4} \varphi^{+a} \varphi^a)^2 - {1
\over 4}
(\varphi^{+a} \varphi^a)^2 + {1 \over 4} \varphi^a \varphi^a \varphi^{+b}
\varphi^{+b}\biggr] \hskip .1cm .\eqno(9)
$$
By an $SU(2) \times U(1)$ transformation we can rotate the vacuum to the
configuration $\varphi^a =\pmatrix{&0\cr&0\cr&v\cr}$ where $v$
is a real quantity. Then the expression of $V_{eff}$ simplifies
$$
V_{ef}(v) = {1 \over G} v^2 - 4 \int {d^4p \over (2\pi)^4} \ell n (p^2 + {1
\over 4}
v^2) \eqno(10)
$$
and the saddle point equation is given by
$$
v\biggl ({1 \over G} - \int {d^4p \over (2\pi)^4}\  {1 \over p^2 + {1 \over 4}
v^2}
\biggr) = 0 \hskip .5cm .\eqno(11)
$$
The energy of the nontrivial extremum is given by
$$
{\cal E}_0 ( <v> )\ =\ 4 \int {d^4 p \over (2\pi )^4}\ \biggl\{ { {1 \over 4}
v^2 \over
p^2 +{1\over 4} v^2} - \ell n \biggl (1 + {v^2 \over 4p^2}\biggr ) \biggr\}
\eqno(12)
$$
which is a negative quantity, as it can be explicitely checked.

With the second set of constraints, considering only the Lorentz scalar degrees
of
freedom in $\phi_2$ and $\phi_3$ defined by $\phi_{2,3}^{\beta \alpha} = {1
\over 4}
\delta^{\beta \alpha} \phi_{2,3}$ the effective potential is
$$\eqalign{
V_{ef} (\phi_1, \phi_2, \phi_3) = &-{4 \over G} \phi^+_1 \phi_1\ -\ {1 \over
2G}
\phi_2 \phi_3\ -\ 2 \int {d^4 p \over (2\pi)^4} \ell n \biggl[ \biggl( p^2 + {1
\over 16}
\phi_2^2 + \phi_1^+ \phi_1\biggr)\cr
&  \times \biggl( p^2 + {1 \over 16} \phi^2_3 + \phi^+_1 \phi_1 \biggr) - {1
\over 16}
\biggl(\phi_2 + \phi_3\biggr)^2 \phi^+_1 \phi_1\biggr]\cr}\eqno(13)
$$
which has the same normalisation as in (9) in the case of zero field
configurations.
With a $U(1)_{1-1}$ tranformation we can make $\phi_1$ real . The saddle point
equations are
$$\eqalignno{
&\phi_1 \Biggl\{ {1 \over G}\ + \int {d^4p \over (2\pi)^4} {p^2 + \phi^2_1 -
{1\over
16} \phi_2 \phi_3 \over \bigl(p^2 + {1 \over 16} \phi^2_2 + \phi^2_1\bigr)
\bigl(p^2 + {1 \over 16} \phi^2_3 + \phi^2_1\bigr) -  {1 \over 16}
\bigl(\phi_2+
\phi_3\bigr)^2 \phi^2_1 }\biggr\} \ =\ 0\hskip .5cm ,\cr
&{2 \over G} \phi_3 + \int {d^4p \over (2\pi)^4} {
\phi_2\bigl (p^2 + {1\over 16} \phi_3^2 + \phi_1^2\bigr) - 2(\phi_2+\phi_3)
\phi_1^2
\over \bigl(p^2 + {1 \over 16} \phi^2_2 + \phi^2_1\bigr) \bigl(p^2 + {1 \over
16}
\phi^2_3 + \phi^2_1\bigr) -  {1 \over 16} \bigl(\phi_2+ \phi_3\bigr)^2 \phi^2_1
} \
=\ 0\hskip .5cm ,&(14)\cr
&{2 \over G} \phi_2 + \int {d^4p \over (2\pi)^4} {
\phi_3\bigl(p^2 + {1\over 16} \phi_2^2 + \phi_1^2\bigr) - 2(\phi_2+\phi_3)
\phi_1^2
\over \bigl(p^2 + {1 \over 16} \phi^2_2 + \phi^2_1\bigr) \bigl(p^2 + {1 \over
16}
\phi^2_3 + \phi^2_1\bigr) -  {1 \over 16} \bigl(\phi_2+ \phi_3\bigr)^2 \phi^2_1
} \
=\ 0\hskip .5cm ,\cr
}$$
which have as the only solutions
$$\eqalignno{
\phi_1 &= 0\hskip .5cm ,\cr
\phi_2 &= -\phi_3 \hskip .5cm ,&(15)\cr
\phi_2 &\biggl ( {2 \over G} - \int {d^4 p \over (2\pi )^4} {1 \over p^2 + {1
\over 16}
\phi^2_2} \biggr ) = 0\hskip .5cm .\cr
}$$
The energy of the nontrivial solution is given by
$$
{\cal E} (< \phi_2 > = < \phi_3 >) = 4 \int {d^4 p \over (2\pi)^4 } \biggl \{
{ {1\over 16} \phi^2_2 \over p^2 + {1 \over 16} \phi^2_2 } - \ell n \biggl ( 1
+
{\phi^2_2 \over 16 p^2 }  \biggr ) \biggr \} \eqno(16)
$$
which is negative.

In order to decide which of the two nontrivial solutions in eqs.(11) and (15)
is the
real minimum, first of all we remark that the vacuum energies (12) and (16)
have
the same functional dependence of the condensates. However in eq.(11) the
critical
coupling is defined by ${1 \over Gc} = \int {d^4 p \over (2\pi)^4} {1 \over
p^2}$ and
in eq.(15) by ${2 \over Gc} = \int {d^4 p \over (2\pi)^4} {1 \over p^2}$. Hence
the
nontrivial solution in (11) will be energetically preferred and the symmetry
$SU(2) \times U(1)_1
\times U(1)_2 \times U(1)_A$ will be broken to $U(1)_{1+2} \times U(1)_A$. In
particular the fermions remain massless.

Comparing the effective potential approach with the DS equations we
remark that we obtain different results. Rewriting the interaction Lagrangian
as
in eq. (6) we find the first term in the right-hand side is repulsive.Its
average on the ground state vanishes because $<{\bar \psi}_1 \psi_2>
=0$. Then it is reasonable to assume that only the second term must be
used in the bouns state analysis. Doing this we find $G_c$ correctly,
the correct vacuum energy and the dependence of the condensate on the
coupling $G$, but the broken symmetries are not the same, as we already
remarked. The second solution from the Dyson-Schwinger equations is not
available in the other
formulation, because it corresponds to auxiliary fields of the type $
{\bar \psi}_1{\tau^a
\over 2} \psi_1$ , $\bar \psi_2 {\tau ^a \over 2} \psi_2$. We cannot rewrite
the
interaction Lagrangian in a simple form using them. More interesting, the real
minimum from the effective potential approach cannot be obtained from the
Dyson-Schwinger eqs. in the one-loop approximation. Namely, trying to write the
eqs. for $S^{(12)}_{ij} = < 0 | T \bar \psi_{1i} \psi_{2j} | 0 >$ we obtain no
nontrivial
solution, related to the fact that we have no $\psi_1 - \psi_2$ mixing at the
tree
level in the Lagrangian.
\vskip 1cm
\noindent {\hbox { \quad {\bf 4. Conclusions }}}
\vskip .5cm
The different results obtained using the two different sets of
constraints indicate that they are not equivalent.  This is a reasonable
result because the two sets have different symmetry transformation
properties and the condensation of one or the other cannot gives the
physics.
   Moreover the real vacuum, corresponding to the breaking of symmetry
$SU(2) \times U(1)_{1-2}$ cannot be obtained in the DS equations
approach in the one-loop approximation, which on the contrary seems to
indicate the breaking of symmetry $U(1)_A$.
   We can ask of possible phenomenological use of such a toy model. The
two isospin doublets $\psi_1$ and $\psi_2$ can be viewed as two quark
families and the symmetries of the model as $SU(2)$ -weak isospin,
$U(1)_1$ -baryon number for the first family, $U(1)_2$ -baryon number
for the second family and $U(1)_A$ -chiral symmetry which perturbatively
forbids the quark masses. The vacuum described by the condensation of
$<{\bar \psi}_1 {\tau \over 2}^a \psi_2>$ spontaneously breaks the weak
isospin and the baryon number down to $U(1)_{1+2}$ which is the total
baryon number. A Cabbibo type mixing is spontaneously generated and the
fermions remain massless.
\vfill
\eject
\noindent {\hbox {\quad {\bf References}}}
\vskip 1cm
\item {[1]} Y. Nambu and G. Jona-Lasinio , Phys.Rev. 122 (1961) 345 .
\item {[2]} Y. Nambu , in Proceedings of the $11^{th}$ Int. Symposium
on Elem.Part.Phys. , Kazimierz , Poland , 1988 ( World Scientific ,
Singapore ,1989 ) , Enrico Fermi Institute Report 89-08 (1989) ;
\item {   } V.A. Miranski , M. Tanabashi and K. Yamawaki , Mod. Phys.
Lett. A4 (1989) 1043 ,  Phys.Lett. B221 (1989) 177 ;
\item {   } W.A. Bardeen , C. Hill and M. Lindner , Phys.Rev. D41 (1990)
1647 .
\item {[3]} D. Gross and A. Neveu , Phys.Rev. D10 (1974) 3235 .
\item {[4]} S. Coleman and E. Weinberg , Phys.Rev. D7 (1973) 1888 ;
\item {   } R. Jackiw , Phys.Rev. D9 (1974) 1686 .

\end